\documentstyle[editedvolume]{crckapb}

\newcommand{\smm}{{submm-mm}}

\begin{opening}
\title{EXTREMELY RED AND DUSTY GALAXIES}
\subtitle{}

\author{P. ANDREANI}
\institute{Dipartimento di Astronomia, Universit\`a di Padova, Italy\\
           MPI f. Extraterrestrische Physik, Garching, Germany}
\author{A. CIMATTI}
\institute{Osservatorio Astronomico di Arcetri, Firenze, Italy}
\author{H. R\"OTTGERING}
\institute{Sterrenwacht, Leiden, The Netherlands}
\author{R. TILANUS}
\institute{JACH, Hawaii, USA and NFRA, Dwingeloo, The Netherlands}

\end{opening}

\runningtitle{EXTREMELY RED GALAXIES}

\begin{document}


\section{ABSTRACT}

Preliminary results of a project aiming at unveiling the nature of
the extremely red galaxies (ERGs) \footnote{\noindent
by extremely red galaxies
hereafter we mean objects with colours $R-K > 6$ and $I-K > 5$} 
found in deep optical-NIR surveys are presented.
Very little is known about these objects, the critical issue being
whether they are old ellipticals at z$>$1 or distant star-forming
galaxies strongly reddened by dust extinction.
\hfill\break
We expect to shed light onto the unknown nature of these galaxies
by completing our three-step project:
(1) the construction of two very deep optical/NIR surveys to select ERGs,
(2) subsequent VLT/NIR spectroscopy; (3)
observations in the \smm ~region with SCUBA at the JCMT and with
MPIfRbolo at the IRAM 30m antenna.

\section{INTRODUCTION}

Optical and NIR deep surveys have recently boosted observational
cosmology and allowed important advances in our understanding
of young galaxies.
It has been possible to detect star-forming objects at high
redshifts and construct the star-formation
rate (SFR) history of the universe converting the detected UV-optical
or line rest-frame luminosities into a star formation rate
(\citeauthor{PIERO} 1996; see also Calzetti, these Proceedings):

Several ways of inferring SFRs have been
used:

\begin{itemize}

\item
The star-formation rate (SFR) in Lyman-break galaxies
deduced from their UV-rest frame luminosity is
of the order of 4-50 h$_{50} ^{-2}$ M$_\odot$/yr (\citeauthor{STEI} 1996;
 \citeauthor{PIERO2} 1998).

\item
Lyman-$\alpha$ emitters present a comparable SFR obtained from their
line luminosity
of 1-10 h$_{50} ^{-2}$ M$_\odot$/yr \cite{HU}

\item OII ($\lambda$3727) emitters have a SFR of 5-60 h$_{50} ^{-2}$ M$_\odot$/yr
\cite{COWIE} 

\item
because of a smaller bias against reddening
higher values for SFR are inferred from H-$\alpha$ and H-$\beta$ line
luminosities: of 5-200 and 10-140 h$_{50} ^{-2}$ M$_\odot$/yr
respectively (\citeauthor{BEC} 1998; \citeauthor{MAN} 1998;
\citeauthor{TEP} 1998; \citeauthor{PET} 1998).
\end{itemize}

When these results are combined together with those from galaxy surveys at
$z \leq 1$
strong constraints on the comoving SFR are inferred \cite{PIERO}.
But a still debated question is the presence of the peak seen at
$z \sim 1.5$, since surveys based only on UV-optical light are strongly
biased towards dust-free objects.

\noindent
Several arguments suggest that the population of high-$z$ galaxies detected
so far may not represent the progenitors of all
local galaxies and the consequent history of SFR may be strongly
underestimated. The presence of other populations of objects is therefore
very likely and may be advocated to explain the following.

\begin{itemize}

\item
There is an absence of passively evolving ellipticals in
deep surveys. Where are the associated protospheroids?

\item
There is a FIR-submm background detected by COBE, very likely
due to the integrated emission of a so-far-hidden population of dusty 
galaxies (\citeauthor{PU96} 1996; \citeauthor{HAUS} 1998).

\item
ISO/SCUBA FIR/\smm ~surveys (\citeauthor{RR} 1997; \citeauthor{TAN} 1997;
\citeauthor{BAR} 1998; \citeauthor{HU98} 1998; \citeauthor{S} 1998;
\citeauthor{ELB}) are key to test the
presence of these dusty objects and are indeed showing a large number of sub-mm
luminous galaxies (at 850$\mu m$ $\sim$ 0.08-0.5 objects arcmin$^{-2}$ with
flux $>$ 3 mJy and 2 objects arcmin$^{-2}$ with $>$ 1mJy, \citeauthor{HU98}
1998).
Is the detected dust obscuring a large fraction of the galaxy 
UV-luminosity?

\end{itemize}

\noindent
{\it Maybe some/all of these dusty objects are already showing up in
OPTICAL/NIR surveys.}

\section{THE QUEST FOR ERGS}

It is important to check whether the new very red galaxies showing up when
we combine optical + NIR images represent part of the population of high-z
dusty galaxies.
These ERGs are {\it missed} by the traditional optical surveys because of
their faintness at optical magnitudes and they do not show up in the
surveys devoted to high-z galaxies such as those mentioned above. They are found
thanks to the combination of deep optical and NIR images both
in random fields and in those containing an AGN. Their surface density in the
field at K$<$20
is of the order of 0.1-0.2 arcmin$^{-2}$ for R-K$>$6 or I-K$>$5 and of $\sim$
0.01-0.05 arcmin$^{-2}$ for R-K$>$7 or I-K$>$6 (\citeauthor{HR} 1994;
\citeauthor{COWIE} 1996; \citeauthor{THOM} 1998; \citeauthor{BAR2} 1998)
(for comparison the surface
density of Lyman-break galaxies is 0.5 arcmin$^{-2}$ while that of
QSOs with B$<$21.5 is ~ 0.015 arcmin$^{-2}$).

If resolved in ground-based and HST images ERGs usually show compact
morphologies. They sometimes have asymmetric and distorted 
morphologies suggesting the presence of an interacting system or 
a tidal arm. Their faintness hampers the exploitation of
Optical/NIR spectroscopy with 4m telescopes
to obtain redshifts and to investigate their nature.
So far the Keck 10m telescope has provided the only 
redshift available of one of these galaxies, HR10 \cite{DEY}.

The existence of a significant population of objects that is
extremely red, moderately bright and at high redshifts
is difficult to explain using the known
properties of nearby galaxies.
It is likely that ERGs form a heterogenous sample with observed properties
alike not because of intrinsic similarity but only because of
selection criteria. It is very unlikely that they are at very
high redshifts (z$>$3), since this would require that these objects
be exotic and very luminous. It is also unlikely that most of them
lie at low redshifts since no population with the properties of
ERGs is yet known to exist locally.
Hints about their nature can be extracted from their extremely red colours
and the likely explanations are actually twofold:

\begin{itemize}

\item
Are they old L$_*$ ellipticals at z$>$1? In this case their red
colours are simply due to the passively evolving population of
stars (Cohen et al., 1998).

\item
Are they strongly extincted starbursts or AGNs? In this case
their UV-optical light is reddened by dust. Are they then similar
to the sub-mm selected galaxies detected from SCUBA?
\end{itemize}

Even if they may not represent a large hidden population of
galaxies in both scenarios they play an important role in
understanding the integrated star-formation in the high-z
population.

\section{OUR ON-GOING PROJECTS}

A multi-wavelength approach was then tackled to unveil the nature
of these objects.
Two surveys are presently being carried out to select two complete
samples of ERGs: one in random fields and one around AGNs
at z$>$1.5.
Their surface abundance will be inferred and targets for VLT
spectroscopy will then be selected.
The surveys make use of ESO ground-based (optical+NIR) and HST
(WFPC2) data (see \citeauthor{POZ} 1998).

Meanwhile, a subsample of the selected ERGs will be observed at
sub-mm and mm wavelengths using SCUBA+JCMT and MPIfRbolo+IRAM. The aim
is to check whether thermal emission from dust in the ISM of these galaxies
can be observed.  The detection of the
dust emission allows also to determine the
FIR luminosity and to infer the SFR. These can be then compared
with the sub-mm selected galaxies. An important outcome of this research
will be then the inference of the ERGs contribution to the global
star-formation history of the Universe and to the FIR/sub-mm
background.

\section{RESULTS OF OUR SUB-MM/MM INVESTIGATION: HR10}

The ERG HR10 (the only one with redshift available) was
independently detected with the IRAM 30m equipped with the 
MPIfRbolo and with the JCMT equipped with the SCUBA double
arrays (\citeauthor{CIM} 1998; see also Dey et al., 1999).
The radio emission of this object is extremely weak in comparison
with the millimetric flux ($\frac{S_{15GHz}}{S_{1mm}} < 0.02 $);
it is very likely therefore that the
detected submm/mm fluxes are not due to synchrotron emission but
to thermal emission from a dusty
medium. Combining these measurements with the ISO upper limit at 175
\,$\mu$m \cite{IVI} one can derive the dust properties.
For a range of dust temperatures between 30 and 45 K the
corresponding total dust mass lies in the range of 8-4 $10^8 h_{50}^{-2}$
M$_\odot$ (for q$_0$=0.5 and a dust emissivity index $\beta$ of 2). One
must note, however, that although the thermal spectrum is not well
constrained due to a lack of data in the Wien region, the uncertainty on
the dust mass is not larger than a factor of 2.
The total rest-frame far-IR luminosity in the range
10-2000\,$\mu$m rest-frame is 2-2.5 10$^{12} h_{50}^{-2}$ L$_\odot$.
This luminosity places
HR10 in the class of ultraluminous infrared galaxies and implies
a SFR (adopting the relationship SFR=$\Psi ~ 10^{-10} L_{FIR} $
and assuming no AGN contribution) of $\sim$ 200-500
h$_{50}^{-2}$ M$_\odot$/yr. It is worthwhile
mentioning here that the SFR deduced from H$\alpha$ emission was
of 80 h$_{50}^{-2}$ M$_\odot$/yr and from the UV continuum of only
1 h$_{50}^{-2}$ M$_\odot$/yr. SFR is then severely underestimated due
to the strong dust extinction.

The nature of this galaxy will be further investigated via interferometric
imaging of the 1.3mm continuum and CO line emission with the Plateau de
Bure IRAM interferometer.

\section{FURTHER SUBMM-MM OBSERVATIONS}

A sample of other 8 ERGs with $K<20$ and $I-K>6$ has been 
observed so far with SCUBA at the JCMT and with the IRAM 30m antenna,
and other observations have been scheduled. 
The final data reduction is currently under way. For 4 ERGs we could 
reach the sensitivity required by our survey (rms$<$2 mJy at 850$\mu$m), 
whereas the weather was not good enough to obtain deep data at 450$\mu$m. 
A preliminary analysis suggests that we obtained three marginal detections at 
850$\mu$m that need to be confirmed with deeper observations. We have also 
searched for the presence of a positive signal from the population of ERGs 
by coadding the data of the whole sample. The weighted average of the 
850$\mu$m flux density of the entire sample provides a signal at $3\sigma$ 
level.
One further object was detected both at 850 and 1250$\mu m$.
Together with the detection of HR10, this hints that 
that at least part of ERGs are dusty, even if not so extreme as HR10.
These findings, however, still need to be confirmed and the final 
results will be presented in a forthcoming paper (Cimatti et al., 
in preparation).
\hfill\break
It should be reminded here that the physical interpretation of the
submm-mm observations is not severely hampered by the fact that the redshifts
of the ERGs are unknown (with the exception of HR10). Thanks
to the strong K-correction caused by the steep grey-body dust 
spectra, the expected flux at $\lambda_{obs}=850\mu$m of a dusty 
star-forming galaxy at $1<z<10$ is not a strong function of the redshift. 
Thus, since ERGs are expected to be at $z>1$, a detection at $\lambda_{obs}
=850\mu$m directly implies a large content of dust and high $L_{FIR}$ and 
SFR irrespective of the redshift (see also \citeauthor{HU98} 1998;
 \citeauthor{BAR} 1998).

\section{CONCLUSIONS}

For at least one galaxy (HR10) a large amount of star formation is
missing from a UV-only census. As such one could consider it as
an observational proof of the predictions by models such as those
by Zepf and Silk (1996), Franceschini et al (1998), Guiderdoni et al. (1998).
One can argue that the space density of
ERGs is not greatly different from those predicted by these models
at these flux levels. For instance,
Guiderdoni et al. (1998) predict a sky surface density of dusty
galaxies at 175 $\mu m$ of 0.05 arcmin$^{-2}$ similar to that of the
extreme red galaxies. 
Sources in the redshift range 0.5-2.5, which contribute to the
cosmic FIR background, should  have fluxes at 200\,$\mu$m (observed)
in the range 10-100 mJy. For T$_d$=18-45 K the expected observed
flux of HR10 at this wavelength would be of 10-40 mJy. 
At this point extrapolation from one object to the entire class
is entirely speculative and a better statistics is required before
carrying out any meaningful comparison.

HR10 can be fully considered a ULIRG since most of its energy is
emitted in the FIR.
\hfill\break
Caution must be used when extrapolating the star
formation rates for UV-selected galaxies to the whole history of SF
in the Universe. At least part of this occurs in highly extincted
environment where UV and optical light cannot escape.
\hfill\break
Objects like HR10 would be missed by optical imaging based on
the continuum break or on strong emission lines, by IRAS and
by traditional quasar surveys.
\hfill\break
Our result demonstrate the powerful tool provided by the
combination of deep optical/NIR imaging with sub-mm/mm
observations.

\end{document}